\documentclass{moriond}

\usepackage{amssymb}

\def\be{\begin{equation}}
\def\ee{\end{equation}}
\def\bea{\begin{eqnarray}}
\def\eea{\end{eqnarray}}

\newcommand{\Photo}{}

\begin{document}
\vspace*{4cm}
\title{Dark matter induced dynamical symmetry breaking}

\author{ K. Kannike }

\address{National Institute of Chemical Physics and Biophysics, \\ Laboratory of High Energy and Computational Physics, R\"{a}vala 10,\\
Tallinn, Estonia}

\maketitle\abstracts{
We consider the classically scale invariant Higgs-dilaton model of dynamical symmetry breaking extended with an extra scalar field that plays the role of dark matter. The Higgs boson is light near a critical boundary between different symmetry breaking phases, where quantum corrections beyond the usual Gildener-Weinberg approximation become relevant. The only large scale, which generates the other scales, is given by the mass of dark matter. This implies a tighter connection between dark matter and Higgs phenomenology. The model has only three free parameters, yet it allows for the observed relic abundance of dark matter while respecting all constraints. The direct detection cross section mediated by the Higgs boson is determined by the dark matter mass alone and is testable at future experiments.}

It is well known that dynamical symmetry breaking together with classical scale symmetry can alleviate the hierarchy problem. If we go further and require that two broken phases nearly coincide, then not only the dilaton but the Higgs boson too will be light. As an example of the general mechanism \cite{Kannike:2021iyh} (model-independent collider phenomenology was discussed in Ref. \cite{Huitu:2022fcw}), we consider a model where quantum corrections that break the symmetry are driven by couplings to scalar singlet dark matter.\cite{Kannike:2022pva} (In the Gildener-Weinberg case, the model was discussed in Refs. \cite{Kang:2020jeg,Ishiwata:2011aa,Gabrielli:2013hma}.) Besides dark matter, this model can provide for cosmic inflation.\cite{Gabrielli:2013hma} Moreover, it is easy to add heavy neutrinos to account for neutrino masses through the type I seesaw mechanism and leptogenesis.

With classical scale invariance, there are no dimensionful terms in the scalar potential. The gauge symmetry of the model will be broken by quantum corrections via the Coleman-Weinberg mechanism.\cite{Coleman:1973jx} At the classical level, there is a continuous flat direction $V = 0$ in the potential. When we take into account quantum corrections, the flat direction is lifted and remains only as a single point. The minimum of the potential lies near the flat direction. In the case of multiple scalars, the Gildener-Weinberg approximation\cite{Gildener:1976ih} tells us that the direction of the minimum is the same as the flat direction. We shall show that in the case of multi-phase criticality, this does not hold.

The tree-level scalar potential of the Higgs boson $h$, dilaton $s$, and dark matter $s'$ of our model is given by
\begin{equation}
  V = \frac{1}{4} \lambda_{H} h^{4} + \frac{1}{4} \lambda_{S}  s^{4} + \frac{1}{4} \lambda_{HS} h^2  s^{2}
  + \frac{1}{4} \lambda_{HS'} h^2 s^{\prime 2}
  + \frac{1}{4} \lambda_{SS'} s^{2} s^{\prime 2}
  + \frac{1}{4} \lambda_{S'} s^{\prime 4}.
\end{equation}
The potential is invariant under a $\mathbb{Z}_2\otimes\mathbb{Z}'_2$ with $s\to - s$ and $s'\to - s'$.

We shall consider the situation in which both the Higgs and the dilaton obtain vacuum expectation values (VEV): $\langle h \rangle \equiv v$ and $\langle s \rangle \equiv w$, respectively.
It is sufficient to require $\lambda_{S'}, \lambda_{HS'}, \lambda_{SS'} > 0$ to avoid a destabilising VEV for dark matter $s'$. Dynamical symmetry breaking can then realise three different phases:
\begin{enumerate}
\item[${h}$)] $\lambda_H =0$ and $\lambda_{S}, \lambda_{HS}>0$ for non-zero $v$,
\item[$s$)] $\lambda_S =0$ and $\lambda_{H}, \lambda_{HS}>0$ for non-zero $w$,
\item[$sh$)] $\lambda_{HS} =-2\sqrt{\lambda_H \lambda_S}< 0$
and $\lambda_{H}, \lambda_{S}>0$ for both $v$ and $w$ non-zero
\end{enumerate}
with the couplings given at the flat direction. The $h$ phase is hard to realise due to the large top Yukawa coupling which gives a negative contribution to the effective potential. The $s$ phase lacks a Higgs VEV and does not break the electroweak symmetry. Thus, only the $sh$ phase has phenomenological appeal.

\begin{figure}
\centerline{\includegraphics[draft=false]{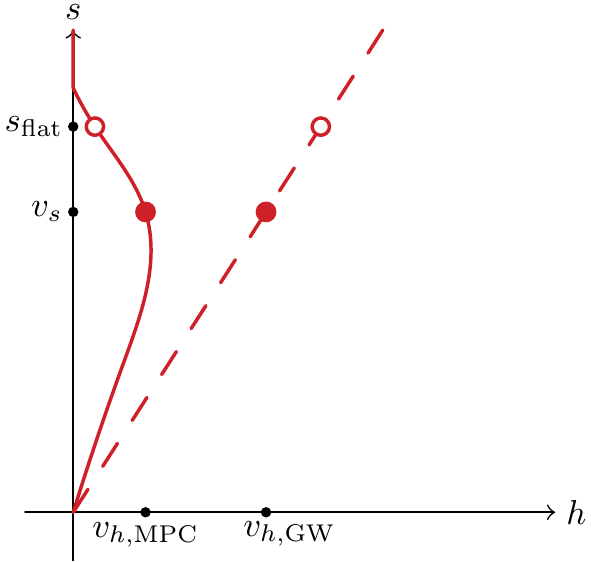}}
\caption{The difference between the Gildener-Weinberg regime (dashed) and the critical regime (solid). The line traces the valley wherein lie the flat direction (empty circle) and the potential minimum (filled circle). Angles are not drawn to scale.}
\label{fig:two:regimes}
\end{figure}

There are two possible regimes for the $sh$ phase. In the usual Gildener-Weinberg regime, $\lambda_{HS}(w) \approx \lambda_{HS}(s_{\rm flat})$ and 
\begin{equation} 
\frac{v}{w} \approx \sqrt{\frac{-\lambda_{HS}(s_{\rm flat})}{2\lambda_H}},
\end{equation}
where $s_{\rm flat}$ is the value of $s$ at the flat direction.
 In the multi-phase critical regime, $\lambda_{S}(s_{\rm flat}) \approx 0$, $\lambda_{HS}(w) \ll \lambda_{HS}(s_{\rm flat}) \approx 0$, the $s$ and $sh$ phases coincide and the Gildener-Weinberg approximation breaks down. The two regimes are illustrated in Fig.~\ref{fig:two:regimes}, where the potential valley wherein lie the flat direction (empty circle) and the minimum (filled circle) is indicated with dashed line in the Gildener-Weinberg case and with solid line in the multi-phase critical regime.
 
In the critical limit, the renormalisation group running is driven by couplings to dark matter. The single dominant mass scale that contributes to the effective potential is the dark matter field-dependent mass
\begin{equation}
m_{s'}^2 = \frac{1}{2} (\lambda_{SS'} s^2 + \lambda_{HS'} h^2).
\end{equation}
We consider the one-loop effective potential
\begin{displaymath}
  V_{\rm eff} = V^{(0)} + V^{(1)} \approx V^{(0)}(\lambda_{i}(\mu = s)),
\end{displaymath}
where we consider a renormalisation group improvement with field-dependent renormalisation scale $\mu = s$. Expanding the remnant of the one-loop part in $h^{2}/s^{2}$, the $\lambda_{S}$ and $\lambda_{HS}$ couplings receive finite shifts from it, but fortunately these can be conveniently absorbed in the $R$ parameter defined below.

In the multi-phase critical regime, the  $\beta$-functions of the Higgs-dilaton portal coupling and dilaton self-coupling are given by 
\begin{equation}
  \beta_{\lambda_{HS}}\approx \frac12 \lambda_{SS'} \lambda_{HS'},\qquad
  \beta_{\lambda_S} \approx \frac14 \lambda_{SS'}^2.
\end{equation}
Notice that the part of $\beta_{\lambda_{HS}}$ proportional to $\lambda_{HS}$ itself that is usually present is neglected here due to the smallness of $\lambda_{HS}$. We see that the dark matter Higgs portal $\lambda_{HS'}$ is important both for symmetry breaking \emph{and} direct detection.
We parametrise the running couplings as
\begin{eqnarray}
\lambda_S^{\rm eff}(s) &=& \frac{{\beta}_{\lambda_S }}{(4\pi)^2}   \ln \frac{s^2}{s^2_S },
\\
\lambda_{HS}^{\rm eff}(s) &=& \frac{{\beta}_{\lambda_{HS} }}{(4\pi)^2} \ln \frac{s^2}{s^2_{HS} }
\\
&=& \frac{{\beta}_{\lambda_{HS} }}{(4\pi)^2} \ln \frac{R \, s^2 }{e^{-1/2}s^2_{S}},
\label{eq:lambda:param}
\end{eqnarray}
where $s_{S}$ and $s_{HS}$ denote the scales where $\lambda_{S}$ and $\lambda_{HS}$, respectively, run through zero, and the ratio $\displaystyle R = e^{-\frac{1}{2}} \frac{s_{S}^{2}}{s_{HS}^{2}}$. If $|\ln R|$ is small, then $s_{HS}$ is close to the flat direction scale and $\lambda_{HS}$ is small at the flat direction: it runs relatively much so we are in the critical regime. For large $|\ln R|$, the $\lambda_{HS}$ is larger at the flat direction and stays almost the same down to the minimum scale, in which case we are in the Gildener-Weinberg regime.

As far as the dilaton is concerned, the symmetry breaking is very close to the single-field Coleman-Weinberg mechanism. The flat direction scale is given by $s_{\rm flat} \approx s_{S}$ and the dilaton VEV $w \approx s_S \, e^{-1/4}$. The Higgs VEV is produced by the deviation of the minimum direction from flat direction due to running,
\begin{equation}
v = \frac{w}{4 \pi} \sqrt{ -\frac{\beta_{\lambda_{HS}} \ln R}{2\lambda_H} }
\end{equation}
and is thus suppressed by quantum corrections. The suppressed Higgs VEV also suppresses the Higgs boson mass. More concretely, the angular deviation is given by
\begin{equation}
  \sqrt{\frac{\lambda_{HS}^{\rm eff}(s_{\rm flat})}{\lambda_{HS}^{\rm eff}(w)}} = \frac{ h_{\rm flat}}{ s_{\rm flat}} {\Bigg /} \frac{v}{w} = 
\sqrt{1 + \frac{1}{2 \ln R}} \quad \mathrm{if} \; \ln R < -\frac{1}{2}.
  \label{eq:flat:over:min}
\end{equation}
We see directly that large $|\ln R|$ corresponds to the Gildener-Weinberg limit.

Because we know that the Higgs mass $m_h \approx \sqrt{2 \lambda_{H}} v \approx 125.1$ GeV and the Higgs VEV $v=246.2$ GeV, the free parameters are $m_{s}$, $m_{s'}$ and $\ln R$. While the dark matter self-coupling $\lambda_{S'}$ is also free, it is largely irrelevant (it should not be too large so as not to bring about a Landau pole near the minimum). Not only the dilaton mass $m^2_{s} \propto \beta_{\lambda_S}$, but also the Higgs mass $m^{2}_{h} \propto \beta_{\lambda_{HS}}$ is loop-suppressed, while the only tree-level mass is $m^{2}_{s'} \propto \lambda_{SS'}$. Inverting the expressions for masses and VEVs, we obtain
\begin{eqnarray}
  \lambda_{SS'} &\approx& \frac{(4 \pi)^{2} m_{s}^{2}}{m_{s'}^{2}},
\\
  \lambda_{HS'} &\approx& -\frac{(4 \pi)^{2} m_{h}^{2}}{m_{s'}^{2} \ln R},
  \\
    w &\approx& \frac{\sqrt{2}{m_{s'}^{2}}}{{4 \pi \, m_{s}}}
\end{eqnarray}
and the Higgs-dilaton mixing angle 
\begin{equation}
\theta\approx \frac{2 \sqrt{2} \pi m_{s} m_{h}^{2} v (1 + \ln R)}{(m_{h}^{2} - m_{s}^{2}) m_{s'}^{2} \ln R}.
\end{equation}

\begin{figure}
\centerline{\includegraphics[draft=false]{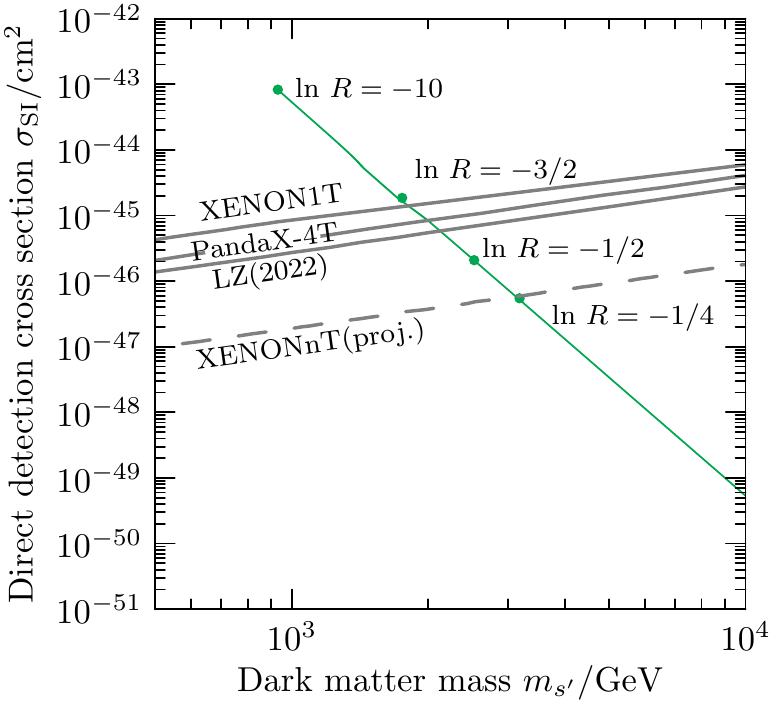}}
\caption{DM direct detection cross section vs. DM mass. Also shown are the strongest limits from direct detection experiments.}
\label{fig:dd}
\end{figure}

The dark matter relic density has to satisfy the observational value $\Omega_{\rm DM} h^{2} = 0.120 \pm 0.001$.\cite{Planck:2018vyg} We can work in the heavy-DM limit $m_{s'}\gg m_s, m_h$, where the non-relativistic DM annihilation cross section for dark matter freeze-out is simply given by
\begin{equation}
\sigma_{\rm ann}  v_{\rm rel} 
\approx\frac{ \lambda_{SS'}^2+4\lambda_{HS'}^2}{64\pi m_{s'}^2}
\approx4\pi^3  \frac{m_s^4+4m_h^4/\ln^2 R}{m_{s'}^6},
\end{equation}
which is Higgs-dominated for $m_{s} \ll m_{h}$ and dilaton-dominated for $m_{s} \gg m_{h}$. The peculiar powers of masses arise from the parametrisation Eq.~\ref{eq:lambda:param}.
 The dark matter effective coupling to nucleons is given by $(f_N m_N/v) \, h\bar N N$,
which yields the spin-independent direct detection cross section
\begin{equation}
\sigma_{\rm SI}\approx
\frac{64 \pi^3 f_N^2 m_N^4}{m_{s'}^6},
\end{equation}
where $m_N=0.946$ GeV is the nucleon mass and $f_N\approx 0.3$. The direct detection cross section is shown in Fig.~\ref{fig:dd} with a green line. For each value of $\ln R$, the line begins at a different dot, corresponding to the Higgs-dominated case. For larger values of the dilaton mass, the coupling $\lambda_{SS'}$ begins to dominate the relic density, so $\lambda_{HS'}$ and therefore the direct detection cross section become smaller. Also shown are the limits from XENON1T (2018),\cite{XENON:2018voc} PandaX-4T (2021),\cite{PandaX-4T:2021bab} LZ (2022),\cite{LZ:2022ufs} and the projected limit from XENONnT.\cite{XENON:2020kmp}

\begin{figure}
\begin{minipage}{0.5\linewidth}
\centerline{\includegraphics[draft=false]{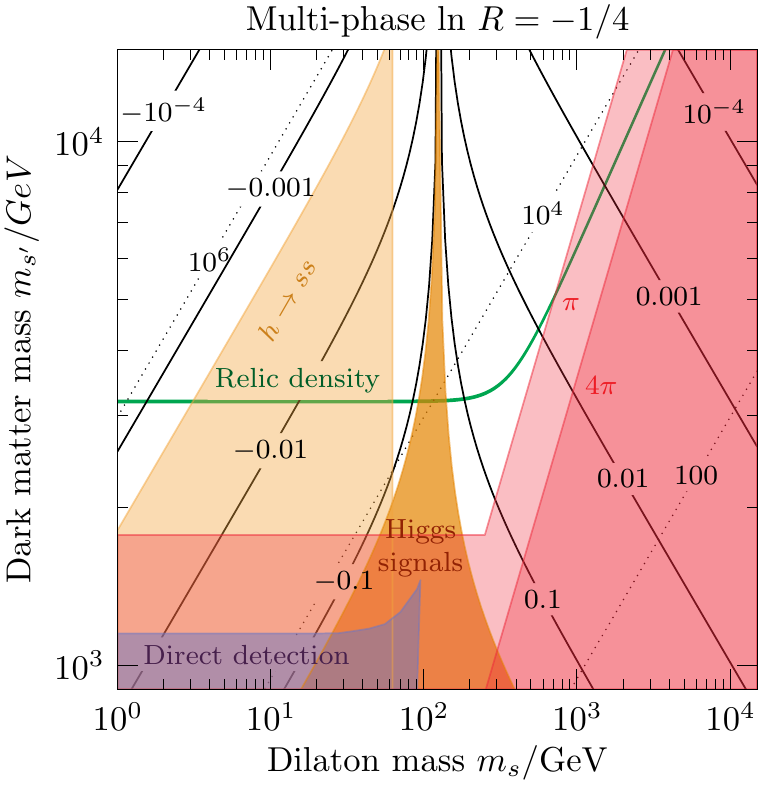}}
\end{minipage}
\hfill
\begin{minipage}{0.5\linewidth}
\centerline{\includegraphics{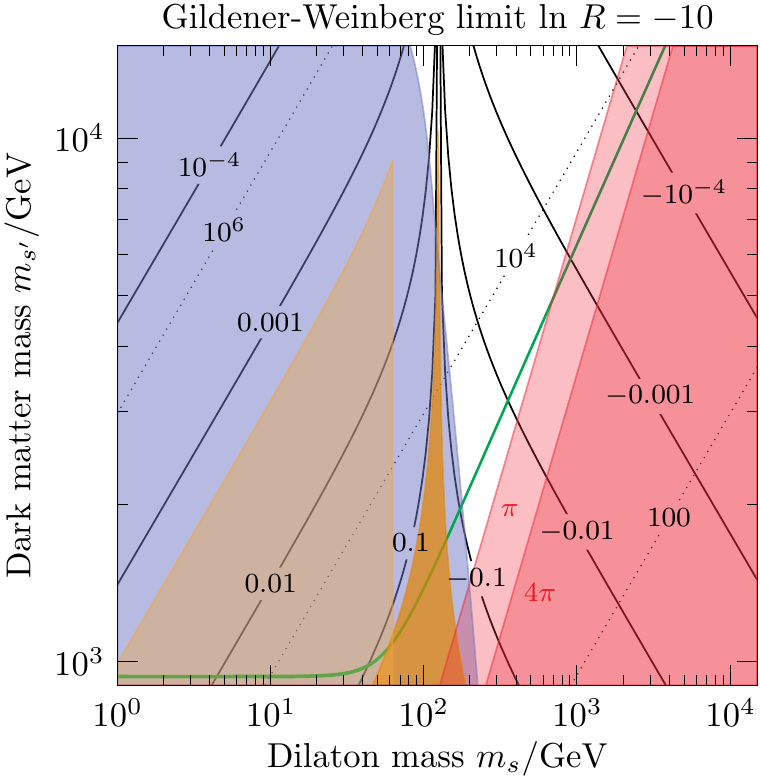}}
\end{minipage}
\caption{Parameter space of the model in the plane of dilaton mass and dark matter mass. Left panel: multi-phase critical case. Right panel: Gildener-the Weinberg limit.}
\label{fig:param:space}
\end{figure}

The full parameter space in the multi-phase critical regime (for $\ln R = -1/4$) and in the Gildener-Weinberg limit (for $\ln R = -10$) is shown in Fig.~\ref{fig:param:space}. The relic density is within the $3\sigma$ Planck bounds inside the green stripe (above the green stripe, dark matter is overabundant). Direct detection constraints are given in lilac, perturbativity bounds in pink, Higgs signals in orange and $h \to s s$ decay constraints in beige.\cite{Robens:2016xkb}

In order to accommodate light neutrino masses, one can easily add the type I seesaw mechanism  with right-handed neutrinos $N_R$.\cite{Minkowski:1977sc,Mohapatra:1979ia,Yanagida:1979as,Gell-Mann:1979vob,Schechter:1980gr} The Yukawa Lagrangian that yields the neutrino mass matrix is given by
\begin{equation}
   -\mathcal{L}_Y = y_{H} \bar{\ell} \tilde{H} N_{R} + \frac{y_{S}}{2}  s \bar{N}^{c}_{R} N_{R} + \rm{h.c.},
\end{equation}
where $\tilde{H} \equiv i \tau_2 H^*$. The heavy neutrinos must be light enough so as to not ruin the mechanism with their negative contributions to the effective potential. With this assumption, the seesaw mechanism, together with leptogenesis, can work on its own.
 
Freeze-in is also possible, in which case the dark matter mass $m_{s'} \sim 10^{8}$~GeV and the scalar couplings are tiny: for dilaton mass $m_{s} = 10$~GeV, we have $\lambda_{SS'} \sim 10^{-11}$, $\lambda_{HS'} \sim 10^{-14}$, $\lambda_{S}\sim 10^{-30}$ and $\lambda_{HS} \sim 10^{-27}$. In the freeze-in scenario, it is practically impossible to detect the model at accessible energies.

In conclusion, there is one large scale in Nature that gives rise to other scales in the critical regime. This scale is provided by dark matter mass and the dynamical symmetry breaking is driven by couplings to dark matter. There are crucial corrections to Gildener-Weinberg approximation in this case, due to which not only the dilaton, but also the Higgs mass is loop-suppressed. Thanks to close relations between symmetry breaking and dark matter, we have a clear prediction for direct detection. For $m_{s} \lesssim m_{h}$, the model can be perturbative up to the Planck scale.

\section*{Acknowledgments}

This work was supported by European Regional Development Fund through the CoE program grant TK133, 
by the Mobilitas Pluss grants MOBTT5, MOBTT86, by the Estonian Research Council grants PRG434, PRG803 and PRG356, and by Italian MIUR under PRIN 2017FMJFMW.

\section*{References}

\bibliography{kristjankannike}

\end{document}